\documentclass[aps,prl,twocolumn,preprintnumbers,floatfix,amsmath,superscriptaddress]{revtex4-1}

\usepackage[bookmarks=false]{hyperref}
\usepackage{graphicx}
\usepackage{amsmath}
\usepackage{xspace}
\usepackage{color}
\usepackage[small]{subfigure}

\RequirePackage[utf8]{inputenc}

\newcommand{\dsigma}{{d}\sigma}

\allowdisplaybreaks[4]

\begin{document}

\preprint{IPPP/18/28, ZU-TH 16/18, CERN-TH-2018-108, MIT-CTP 5013, HU-EP-18/14}


\title{Precise QCD Description of the Higgs Boson Transverse Momentum Spectrum}

\author{Xuan~Chen}
\affiliation{Department of Physics, University of Z\"{u}rich, CH-8057 Z\"{u}rich, Switzerland}
\email{xuan.chen@uzh.ch, thomas.gehrmann@uzh.ch}
\author{Thomas~Gehrmann}
\affiliation{Department of Physics, University of Z\"{u}rich, CH-8057 Z\"{u}rich, Switzerland}
\author{E.W.~Nigel Glover}
\affiliation{Institute for Particle Physics Phenomenology, Department of Physics, University of Durham,
	Durham, DH1 3LE, UK}
\email{e.w.n.glover@durham.ac.uk}
\author{Alexander~Huss}
\affiliation{Theoretical Physics Department, CERN, CH-1211 Geneva 23, Switzerland}
\email{alexander.huss@cern.ch}
\author{Ye~Li}
\affiliation{Fermilab, PO Box 500, Batavia, IL 60510, USA}
\email{yli32edu@gmail.com}
\author{Duff~Neill}
\affiliation{Theoretical Division, MS B283, Los Alamos National Laboratory, Los Alamos, NM 87545, USA}
\email{duff.neill@gmail.com}
\author{Markus~Schulze}
\affiliation{Institut f\"{u}r Physik, Humboldt-Universit\"{a}t zu Berlin, D-12489 Berlin, Germany}
\email{markus.schulze@physik.hu-berlin.de}
\author{Iain~W.~Stewart}
\affiliation{Center for Theoretical Physics, Massachusetts Institute of Technology,
	Cambridge, MA 02139, USA}
\email{iains@mit.edu}
\author{Hua~Xing~Zhu}
\affiliation{Department of Physics, Zhejiang University, Hangzhou, 310027, China}
\email{zhuhx@zju.edu.cn}

\begin{abstract}
The transverse momentum ($p_T$) distribution of Higgs bosons at hadron colliders enables a detailed probe of its production dynamics and is a key ingredient to precision studies of Higgs boson properties, but receives very large QCD corrections. We obtain a precision prediction for the $p_T$ spectrum by matching second-order (NNLO) QCD corrections at large $p_T$ with resummation of third-order logarithmic (N$^3$LL) corrections at small $p_T$.  We achieve significantly improved results for $p_T< 35\,{\rm GeV}$ with perturbative uncertainties $\lesssim \pm 6\%$, and thus a convergent perturbative series for all values of $p_T$. 
\end{abstract}

%
%

\maketitle

\newpage

\textit{Introduction.---}
\label{sec:intro}
With the increased statistics and excellent performance of the experiments at the Large Hadron Collider (LHC), precision analysis of the production and decay of the Higgs boson \cite{Chatrchyan:2012xdj,Aad:2012tfa,Aad:2014lwa,Khachatryan:2015rxa}, the linchpin of the Standard Model, is now becoming reality. Concurrent with the experimental advances has been a concerted effort to improve the theoretical calculations of Higgs production through the gluon fusion mechanism ~\cite{deFlorian:1999zd,Harlander:2002wh,Anastasiou:2002yz,Ravindran:2003um,Ravindran:2002dc,Anastasiou:2015ema,Mistlberger:2018etf,Bozzi:2005wk,Mantry:2009qz,deFlorian:2011xf,Becher:2012yn,Neill:2015roa,Boughezal:2015dra,Boughezal:2015aha,Caola:2015wna,Chen:2016zka,Caola:2016upw,Frederix:2016cnl,Jones:2018hbb,Vladimirov:2016dll,Monni:2016ktx,Bizon:2017rah,Caola:2018zye}, driven by the large corrections one finds in the first few orders of perturbation theory. A key observable is the transverse momentum ($p_T$) spectrum of the Higgs boson with respect to the beam directions, which quantifies how the Higgs boson recoils against partonic radiation, thereby probing the interaction of the Higgs boson with Standard Model particles and possible new states. Calculating the full spectrum requires a combination of fixed order perturbation theory in $\alpha_s$ and a resummation to all orders in $\alpha_s$  of the most singular large logarithms $\ln(p_T/m_H)$ for  $p_T\ll m_H$, where $m_H$ is the Higgs mass.  In this letter, we compute the transverse momentum spectrum under the infinite top quark mass assumption at the highest precision currently possible, by combining next-to-next-to-leading order (NNLO) at large transverse momentum~\cite{Chen:2016zka} with an all-order resummation (based on soft-collinear effective field theory, SCET~\cite{Bauer:2000ew,Bauer:2000yr,Bauer:2001yt,Bauer:2002nz}) of large logarithmic corrections at small transverse momentum, including sub-leading logarithms up to the third level (N$^3$LL) using the recently computed 3-loop rapidity anomalous dimension~\cite{Li:2016axz,Li:2016ctv}. 
Our numerical NNLO results extend to considerably smaller values of 
 $p_T$ compared to earlier work~\cite{Boughezal:2015dra,Chen:2016zka}. 
Resummation at N$^3$LL was previously achieved in a momentum space resummation using the \texttt{RadISH} Monte Carlo program~\cite{Bizon:2017rah} and matched to 
NNLO results at low resolution~\cite{Boughezal:2015dra}. Our approach to obtain N$^3$LL resummation uses a SCET factorization formula, which yields analytic expressions for all singular fixed-order terms. Taken together, these two advancements enable a detailed cross-validation of the results from fixed-order and resummation in all parton-level channels, and yield the first high-resolution description of the Higgs boson $p_T$ spectrum. 


\textit{Method.---}
The dominant Higgs boson production process at the LHC is gluon fusion, mediated by a top quark loop. This process 
can be described by integrating out the top quark, resulting in 
an effective field theory (EFT) coupling the Higgs boson to  
the gluon field strength tensor~\cite{Wilczek:1977zn,Shifman:1978zn,Inami:1982xt}, combined 
with QCD containing five massless quark flavors. The Wilson coefficient of this effective 
operator is known to three loop accuracy \cite{Chetyrkin:1997un}, and the validity of this  description up to 
$p_T$ of the order of the top quark mass has been established~\cite{Frederix:2016cnl,Neumann:2016dny} in detail. For higher 
$p_T$ the exact top quark mass dependence needs to be 
included, and a recent NLO QCD calculation~\cite{Jones:2018hbb} 
has demonstrated that this can be accounted for by a multiplicative rescaling of the EFT predictions. The results 
presented here focus on $p_T$ below the top quark threshold, and are obtained in the EFT framework. 

The $p_T$ distribution of the Higgs boson, valid at both small and large $p_T$, can be written as
\begin{align} \label{eq:sigmatotal}
\frac{d\sigma}{dp_T^2}
 &=\frac{d\sigma^{\rm r}}{dp_T^2} + \left( \frac{d\sigma^{\rm f}}{dp_T^2}-\frac{d\sigma^{\rm s}}{dp_T^2} \right)
   \,,
\end{align}
where $d \sigma^{\rm r}/dp_T^2$ is the resummed distribution, $d\sigma^{\rm f}/dp_T^2$ is the fixed order distribution, and $d\sigma^{\rm s}/d p_T^2$ is the fixed-order singular distribution.
The resummed distribution $d \sigma^{\rm r}/dp_T^2$ can be written as an inverse Fourier transformation from impact parameter space $\vec{b}$ to momentum space $\vec{p}_T$, as given in Eq.~\eqref{eq:resum_xsec}.
 Here, $\sigma_0$ is the Born cross section for $gg\to H$ and we denote $|\vec{b}| = b$ and $b_0 = 2 e^{-\gamma_E}$. 
\begin{widetext}
	\begin{gather}
	\label{eq:resum_xsec}
	\frac{d\sigma^{\rm r}}{dp_T^2}= \pi\sigma_0\int dx_a dx_b\delta\Big(x_ax_b-\frac{m_H^2}{E_{\rm CM}^2}\Big) \int\frac{d^2\vec{b}}{(2\pi)^2}e^{i\vec{p}_T\cdot \vec{b}} W(x_a, x_b, m_H, \vec{b})  \,,
	\\
	\label{eq:W}
	W(x_a, x_b, m_H, \vec{b}) = H(
	m_H, \mu_h)  U_h(m_H,\mu_B, \mu_h)   S_\perp( \vec{b} , \mu_s, \nu_s
	)  U_s(b,\mu_B, \mu_s; \nu_B, \nu_s)  \prod_{\gamma = a, b}B^{\alpha\beta}_{g/N_\gamma}( x_\gamma,
	\vec{b}, m_H,  \mu_B, \nu_B ) \,,
	\\
	U_h(m_H,\mu, \mu_h) =   \exp\left[ 
	2 \int^{\mu}_{\mu_h} \frac{d \bar{\mu}}{\bar{\mu}}
	\left( \Gamma_{\rm cusp}  \big[ \alpha_s(\bar{\mu}) \big]  \ln \frac{m_H^2}{\bar{\mu}^2} +
	\gamma_V \big[ \alpha_s (\bar{\mu}) \big] \right) \right] \,,
	\label{eq:Uh}
	\\
	U_s(b,\mu, \mu_s; \nu, \nu_s) = 
	\exp\left[
	2 \int_{\mu_s}^\mu \frac{d\bar{\mu}}{\bar{\mu}} \left( 
	\Gamma_{\rm cusp} \big[ \alpha_s(\bar{\mu}) \big] \ln \frac{b^2 \bar{\mu}^2}{b_0^2} - \gamma_s \big[\alpha_s(\bar \mu) \big] 
	\right)
	\right] \left( \frac{\nu^2}{\nu_s^2}\right)^{ \int\limits_\mu^{b_0/b} \frac{d\bar \mu}{\bar \mu} 2  \Gamma_{\rm cusp}\big[\alpha_s(\bar \mu) \big] + \gamma_r \big[ \alpha_s(b_0/b)\big] } \,,
	\label{eq:Us}
	\\
	B_{g/N}^{\alpha\beta} (x, \vec{b}, m_H, \mu, \nu) = \sum_j \int_x^1 \frac{d z}{z} \left[
	\frac{g^{\alpha\beta}_\perp}{2} I_{gj}(z,\vec b, m_H, \mu, \nu) +
	\left( \frac{g_\perp^{\alpha\beta}}{2} + \frac{b^\alpha b^\beta}{b^2} \right) I_{gj}'(z,\vec b, m_H, \mu, \nu) 
	\right] f_{j/N} (x/z, \mu) \,.
	\label{eq:beam}
	\end{gather}
\end{widetext}
In Eqs.~\eqref{eq:resum_xsec} and \eqref{eq:W}, the kernel of the integral, $W$, has been factorized into products of a hard function $H(m_H, \mu_h)$, a soft function $S_\perp (\vec b, \mu_s, \nu_s)$, and the beam functions $B_{g/N_\gamma}^{\alpha\beta} (x_\gamma, \vec b, m_H, \mu_B, \nu_B)$, each evaluated at an appropriate scale to avoid large logarithms, and subsequently evolved to common scales. 
Here $f_{j/N}$ are the standard $\overline {\rm MS}$ parton distribution functions and $I_{gj}$ and $I_{gj}'$ are perturbatively calculable matching coefficients. Note that besides the usual renormalization scale $\mu$, the soft function and beam function also depend on the rapidity scale $\nu$.
For $p_T\ll m_H$ the resummation is carried out by making the canonical scale choices 
\begin{gather}
\label{eq:canonical}
\mu_h = m_H\,, \quad \mu_B = \mu_s = \nu_s = b_0/b \,, \quad \nu_B=m_H\,,
\end{gather}
which ensures there are no large logarithms in $H$, $S_\perp$ and $B_{g/N_\gamma}^{\alpha\beta}$ in Eq.~(\ref{eq:W}). Large logarithms are then resummed through the evolution factors $U_h(m_H,\mu_B, \mu_h)$ and $U_s(b,\mu_B, \mu_s; \nu_B, \nu_s)$, which connect the hard scale and the soft scales to the scales of the beam function, respectively. The $U_h$ and $U_s$ are derived from SCET renormalization group and rapidity renormalization group equations~\cite{Chiu:2011qc,Chiu:2012ir}.
They depend on the cusp anomalous dimension $\Gamma_{\rm cusp}$~\cite{Moch:2004pa}, the gluon anomalous dimension $\gamma_V$~\cite{Baikov:2009bg,Lee:2010cga,Gehrmann:2010ue}, the soft anomalous dimension $\gamma_s$~\cite{Becher:2007ty,Li:2014afw}, and the rapidity anomalous dimension $\gamma_r$~\cite{Li:2016ctv,Vladimirov:2016dll}, each of which is now known to three loops. In the case of the cusp anomalous dimension, the four-loop leading color approximation is also known~\cite{Moch:2017uml}. The initial condition of evolution for the hard function and soft function are also known to three-loop order~\cite{Baikov:2009bg,Lee:2010cga,Gehrmann:2010ue,Li:2014afw}. In the case of the beam function in Eq.~\eqref{eq:beam}, the initial conditions $I_{gj}$ and $I_{gj}'$ are known to two loops~\cite{Gehrmann:2014yya,Echevarria:2016scs}, and the logarithmic terms are known to three loops, which involves the 3-loop splitting functions~\cite{Moch:2004pa,Vogt:2004mw}.  Our N$^k$LL resummed calculation is obtained from these ingredients, where we include  all logarithmic terms at ${\cal O}(\alpha_s^k)$ in the $H$, $S_\perp$ and $B_{g/N_\gamma}^{\alpha\beta}$ boundary conditions. For larger $p_T\sim m_H$ the resummation has to be turned off so that $d\sigma^{\rm r}/dp_T^2 = d\sigma^{\rm s}/dp_T^2$ in Eq.~(\ref{eq:sigmatotal}) and the cross section reduces to the fixed order result. This is achieved by making a transition of the various $\mu_i$ to a single scale, and also transitioning to a single rapidity scale, using profile functions~\cite{Ligeti:2008ac,Abbate:2010xh} as explained below.

\begin{figure}
\includegraphics[scale=0.65]{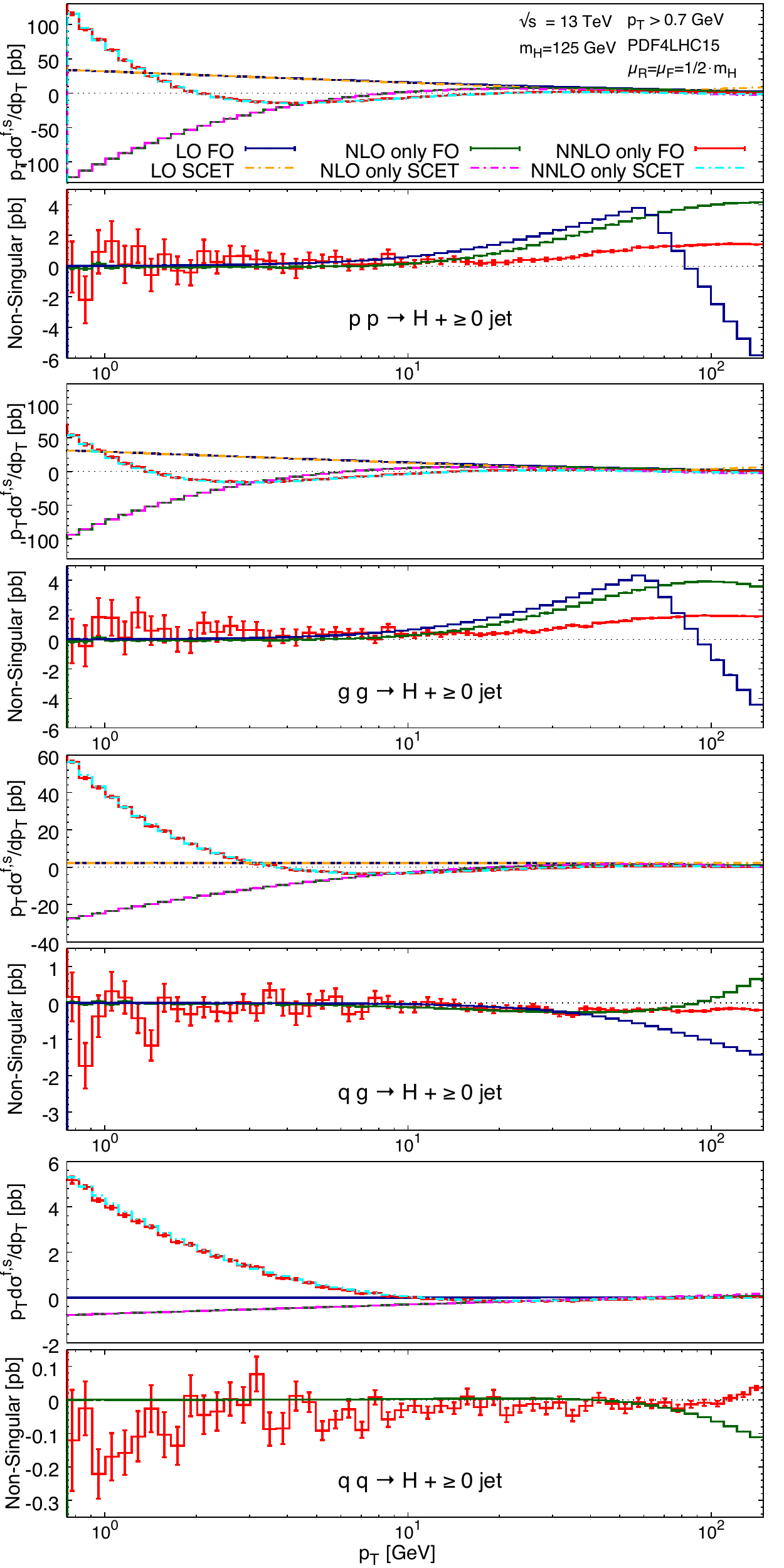}
\caption{\label{fig:compare} Comparison of the transverse momentum spectrum between fixed-order perturbation theory (FO) and singular terms from the expansion of the resummed prediction (SCET), using the sum of all partonic channels ($pp$) or with individual partonic channels ($gg$, $qg$, $qq$). In individual channels, $q$ denotes the sum of quark and anti-quark of all flavors. }
\end{figure}
The fixed-order (FO) distribution $d \sigma^{\rm f}/d p_T^2$ is obtained as a perturbative expansion in 
 $\alpha_s$ by combining all parton-level contributions that contribute to a given order. As the transverse momentum is fixed to a non-vanishing value, the leading order process is given by the production of the Higgs boson recoiling against a parton (As a result, the counting of orders (NNLO: ${\cal O}({\alpha_s^3})$) is performed with respect to $p_T$ at finite values. 
For the inclusive Higgs boson production, these would constitute corrections at N$^3$LO.).
  At higher orders in perturbation theory, a method for the 
 combination of infrared-singular real radiation and virtual loop corrections becomes necessary. In this letter, we employ the antenna subtraction method~\cite{GehrmannDeRidder:2005hi,Daleo:2006xa,Currie:2013vh} to compute the fixed-order distribution up to NNLO
 in perturbation theory. The calculation is performed within the parton-level event generator {\tt NNLOJET}~\cite{Chen:2016zka}, which combines the tree-level double real radiation corrections (RR,~\cite{DelDuca:2004wt,Dixon:2004za,Badger:2004ty}), 
the one-loop single real 
radiation corrections (RV,~\cite{Dixon:2009uk,Badger:2009hw,Badger:2009vh}) and 
 the two-loop virtual corrections (VV,~\cite{Gehrmann:2011aa}) 
with appropriate antenna subtraction terms. Distributions are obtained as binned histograms in $p_T$. 
Schematically, the NNLO contribution to the distribution takes the 
form 
\begin{eqnarray}
\frac{\dsigma_{NNLO}}{dp_T^2}&=&\int_{\Phi_{3}}\left[\frac{\dsigma_{NNLO}^{RR}}{dp_T^2}
-\frac{\dsigma_{NNLO}^S}{dp_T^2}\right]
\nonumber \\
&+& \int_{\Phi_{2}}
\left[
\frac{\dsigma_{NNLO}^{RV}}{dp_T^2}-\frac{\dsigma_{NNLO}^{T}}{dp_T^2}
\right] \nonumber \\
&+&\int_{\Phi_{1}}\left[
\frac{\dsigma_{NNLO}^{VV}}{dp_T^2}-\frac{\dsigma_{NNLO}^{U}}{dp_T^2}\right],
\label{eq:fo}
\end{eqnarray}
where the antenna subtraction terms $\dsigma_{NNLO}^{S,T,U}$ (which upon integration add up to zero) 
ensure the finiteness of each $n$-parton phase space 
integral $\int_{\Phi_n}$. Details of the calculation are described in~\cite{Chen:2016zka}, where the 
transverse momentum distribution was computed to NNLO for large $p_T$. Extending the lower bound on $p_T$ towards smaller values becomes increasingly challenging due to the large dynamical range probed in the phase-space integration and the associated numerical instabilities.  We adapted the {\tt NNLOJET} code 
to cope with this task and further split the integration region into several intervals in 
 $p_T$ and applied dedicated reweighting factors in each region. With these optimizations, fixed-order predictions are obtained down to  $p_T = 0.7$~GeV,  both for a linear binning of 1~GeV width and a logarithmic spacing  of ten bins per 
 ${e}$-fold. 

At small transverse momentum $p_T$ with respect to the Higgs mass $m_H$, the cross-section
can be split into a singular (s) and non-singular (n) piece: 
\begin{equation}
\label{eq:fixed_order_small_pt}
\frac{d\sigma^{\rm f}}{d p_T^2} =  \frac{d \sigma^{\rm s}}{d p_T^2} + \frac{d \sigma^{\rm n}}{d p_T^2} 
\end{equation}
with
\begin{eqnarray}
\frac{d \sigma^{\rm s}}{d p_T^2} & =& 
\frac{ \sigma_0}{p_T^2} \sum_{i=1}^{\infty} \left(\frac{\alpha_s}{\pi} \right)^i  \sum_{j=0}^{2 i - 1}  c_{i,j}
 \ln^j \frac{p_T^2}{m_H^2} \,,
\label{eq:sigs} \\
\frac{d \sigma^{\rm n}}{d p_T^2} & =&  \mathcal{O}\left((p_T/m_H)^0\right)\,.
\end{eqnarray}
The coefficients $c_{i,j}$ are obtained analytically (up to the integrals over the PDFs) by setting the evolution factors $U_h$ and $U_s$ to unity in Eq.~(\ref{eq:W}), and evaluating the hard function,  soft function, and beam function at common scales $\mu_i=\mu_F=\mu_R$ and $\nu_s=\nu_B$.  To compare these singular terms with the full fixed order prediction, we integrate Eq.~\eqref{eq:sigs} over the same binnings that were used in the numerical evaluation in Eq.~\eqref{eq:fo}. Obtaining terms up to the single logarithms  in Eq.~\eqref{eq:sigs} requires using: NLL at LO, NNLL at NLO and N$^3$LL at NNLO.

For the numerical results, we use the PDF4LHC15 (NNLO) PDFs \cite{Butterworth:2015oua} from the LHAPDF library \cite{Buckley:2014ana} with its central value of $\alpha_s(m_Z)=0.118$. The center of mass energy of pp collisions is set to 13 TeV. The mass of the Higgs boson and top quark (dependence in the Wilson coefficient) are set to be $m_H=125$ GeV and $m_t=173.2$ GeV. The central values for the factorization and renormalization scales are chosen as $\mu_F=\mu_R=m_H/2$, with the theory error from fixed order calculations estimated from the envelope of a three-point variation between $m_H/4$ and $m_H$.

Figure~\ref{fig:compare} compares the  fixed-order contributions at LO, NLO and NNLO for the transverse momentum spectrum where the curve labeled as the SCET prediction is $p_T d\sigma^{\rm s}/dp_T = 2 p_T^2\, d\sigma^{\rm s}/dp_T^2$ at the corresponding order. For better visibility, the distributions are multiplied by $p_T$, and each higher order contribution is displayed separately, instead of being added 
to the previous orders. The bottom panels show the difference of the two curves, i.e.~the non-singular parts which should behave as $p_T d \sigma^{\rm n}/dp_T \sim {\cal O}(p_T^2)$ for $p_T\ll m_H$, to further elucidate the low-$p_T$ behavior between the two predictions. The top frame shows that the small $p_T$ behavior of the fixed-order 
spectrum is in excellent agreement with the predicted singular terms in Eq.~(\ref{eq:sigs}). This agreement is further substantiated
in the lower three frames, where individual parton-level initial states are compared (with $q$ denoting the 
sum over quarks and antiquarks of all light flavors). We point out that the 
(numerically subdominant) $qq$ channel turns out to be the numerically most challenging, since contributions 
from valence-valence scattering favor events with higher partonic center-of-mass energy than in any of the 
other channels. The excellent agreement between fixed-order perturbation theory and SCET-predictions for the singular 
terms serves as a very strong mutual cross check of both approaches. It demonstrates that our calculation of the non-singular terms 
is reliable over a broad range in $p_T$, thereby enabling a 
consistent matching of the NNLO and N$^3$LL predictions.  
%
\begin{figure}
\includegraphics[scale=0.75]{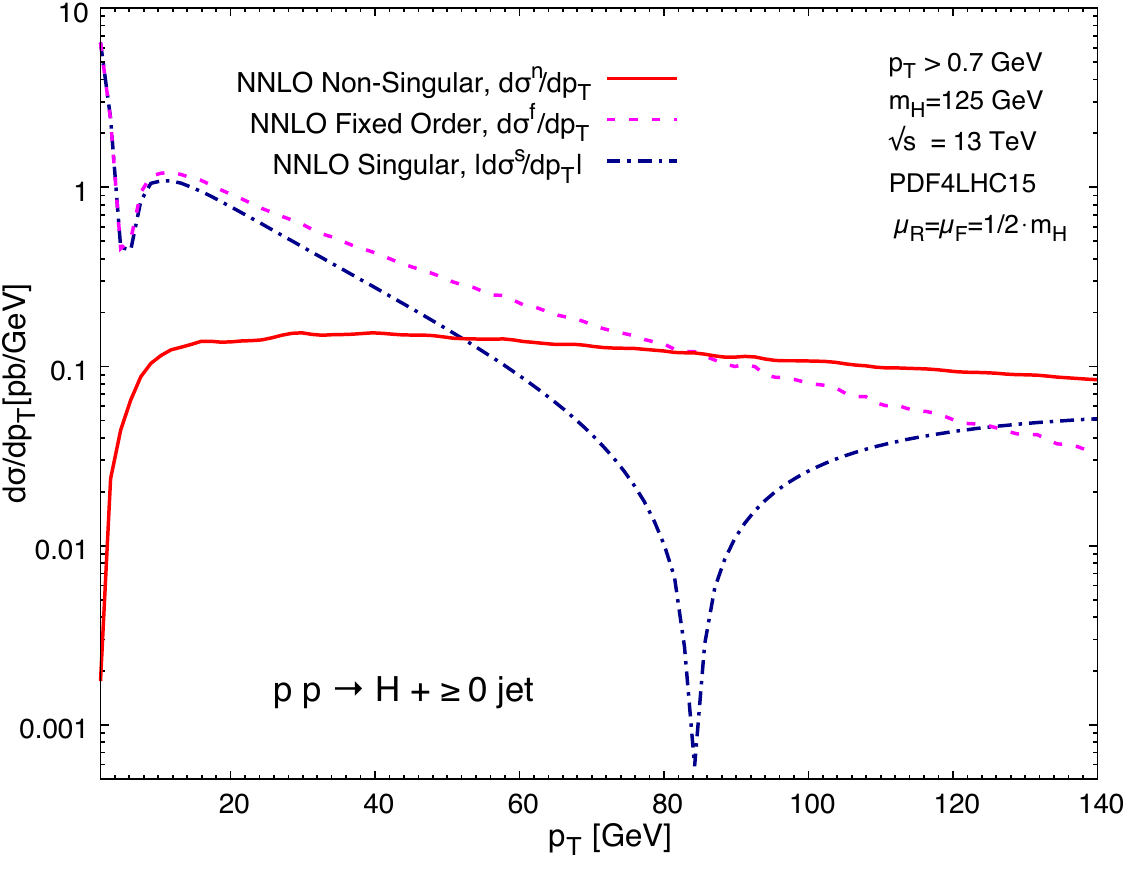}
\caption{\label{fig:logy_compare} Comparison of full fixed-order spectrum, the absolute value of singular distribution, and the non-singular distribution through to NNLO. Here $d\sigma^{\rm n}/dp_T \sim {\cal O}(p_T)$ for $p_T\ll m_H$.}
\end{figure}

\textit{Matching and results.---} 
For a reliable description of the transverse-momentum spectrum, the resummation of large logarithms in $d\sigma^{\rm s}/d p_T^2$ has to be turned off at large $p_T$. This can be seen clearly from Fig.~\ref{fig:logy_compare}, which depicts the full fixed-order spectrum, the absolute value of singular distribution, and the non-singular distribution, all through to NNLO. At $p_T \ll 50$ GeV, the singular distribution dominates the fixed-order cross section, and the resummation of higher order logarithms is necessary. Around $50$ GeV, the singular and non-singular distribution become comparable, and resummation has to be gradually turned off.
The NNLO singular pieces turn negative and largely constant above $pT\approx 85$ GeV, indicating that the decomposition of the NNLO fixed order prediction into singular and non-singular pieces is losing its meaning there.
There are several different prescriptions on how to turn off the resummation~\cite{Collins:1984kg,Arnold:1990yk,Balazs:1997xd,Bozzi:2005wk,Boer:2014tka,Neill:2015roa,Collins:2016hqq,Bizon:2017rah}. In this letter, we follow Ref.~\cite{Neill:2015roa} by introducing $b$ and $p_T$ dependent profile functions, defining
\begin{align}
\rho(b,p_T)&=\frac{\rho_{l}}{2}\Big[1-\text{tanh}\Big(4s\Big(\frac{p_T}{t}-1\Big)\Big)\Big]\nonumber\\
&\qquad+\frac{\rho_{r}}{2} \Big[1+\text{tanh}\Big(4s\Big(\frac{p_T}{t}-1\Big)\Big)\Big] \,,
\end{align}
where $\rho(b,p_T)$ is used for $\mu_s=\mu_s(b,p_T)=\mu_B$,  $\nu_s=\nu_s(b,p_T)$, and $\mu_h=\mu_h(p_T)$, which appear in Eq.~\eqref{eq:W}. 
$\rho_l$ is the initial scale for each profile, taken to be the canonical scales in Eq.~\eqref{eq:canonical} so that at small $p_T$ the large logarithms are resummed. $\rho_r$ is the final scale for each profile, which is chosen to be $\mu_h=\mu_B=\mu_s=\mu_F=\mu_R$, while for $\nu_s$ it is $m_H$. 
The parameters $s$ and $t$ govern the rate of transition between the fixed order result and the resummation, where the transition starts at $p_T\simeq t - t/(2s)$, is centered at $p_T=t$, and ends at $p_T\simeq t + t/(2s)$. In our calculation, we choose $s=1$, and $t = 20,\ 25,\ 30,\ 35,\ 40,\ 50$ GeV to estimate the uncertainties from different profiles. 
%
The uncertainties for the final resummed + fixed-order prediction are estimated by three-point variations of i)  the $\rho_l$ for $\mu_h$ about $m_H$ and $\rho_r$ for all scales (varied simultaneously), and ii) the $\rho_l$ for $\mu_B=\mu_s$ and $\nu_s$ about $b_0/b$ (varied independently). We always fix $\nu_B=m_H$. We take the envelope of the resulting 66 curves as the uncertainty band at each order. 
Further uncertainties in our calculation include the missing four-loop cusp anomalous dimension and the treatment of non-perturbative corrections at large $b$. They are estimated to be negligible compared with the aforementioned scale uncertainties. Additional independent uncertainties related to the parton distributions and value of $\alpha_s(m_Z)$ should be included for a detailed phenomenological study.

\begin{figure}
	\includegraphics[scale=0.73]{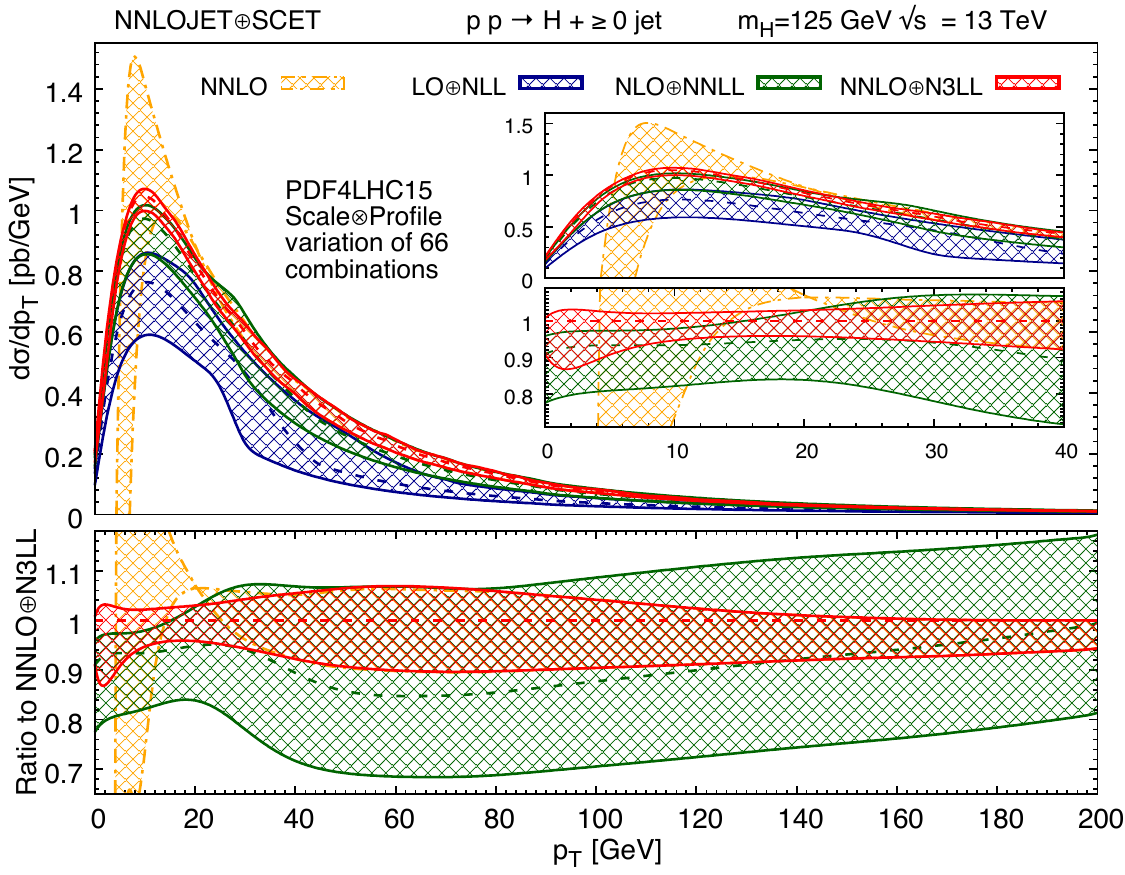}
	\caption{\label{fig:money_plot} The Higgs-boson transverse momentum distribution matched between FO and SCET. 
		Dashed lines indicate central scales of $m_H/2$ and matching profile centered at 30~GeV. 
		The theoretical uncertainties are estimated by taking the envelope of all scale and profile variations (see text). Ratio plots in the lower panel presents the scale and profile variation with respect to NNLO+N$^3$LL (red dashed line). }
\end{figure}

The final matched transverse momentum spectrum is shown in Fig. \ref{fig:money_plot}. We plot the distributions at LO+NLL, NLO+NNLL, and NNLO+N$^3$LL.  We also plot the unmatched NNLO distribution. At small transverse momentum, the fixed order distribution displays unphysical behavior, due to the presence of large logarithms. We see that the matched distribution smoothly merges into the fixed-order cross section around 40 GeV, and that the scale uncertainty reduces order-by-order in perturbation theory. The perturbative uncertainties at NNLO+N$^3$LL have been reduced to $\lesssim \pm 6\%$ for $5 < p_T< 35\,{\rm GeV}$, are $\pm 10\%$ for intermediate $p_T$, and decreasing again at large $p_T$.


\textit{Conclusions.---}
In this letter we have presented for the first time precise predictions for the Higgs transverse momentum spectrum at small $p_T$, with resummation at N$^3$LL matched to fixed-order results at NNLO. The calculation builds upon efficient subtraction formalism for jet processes, improved formalism for resummation of large transverse logarithms, and known high-order anomalous dimensions and matching coefficients. We use an additive matching scheme which relies on the extraction of non-singular corrections from singular ones, and usually requires numerical Monte Carlo precision at the level of 1 per-mille, which imposes a strong challenge on fixed-order calculations in the infrared unstable small $p_T$ region. We have shown excellent agreement between SCET  and {\tt NNLOJET} in this region, which provides a highly nontrivial check of both calculations. The final matched predictions show a continuous reduction of scale uncertainties order by order, and are significantly more precise for small $p_T$. We expect our results will have an important impact on understanding the detailed properties of the Higgs boson at the LHC. 

\textit{Acknowledgements.---}
We  thank the University of Zurich S3IT and CSCS Lugano for providing the computational resources for this project.
This research was supported in part by the UK Science and Technology Facilities Council under contract ST/G000905/1, by the
Swiss National Science Foundation (SNF) under contracts 200020-175595 and
CRSII2-160814, by the Swiss National Supercomputing Centre (CSCS) under
project ID UZH10, by the Research Executive Agency (REA) of the European Union under the
ERC Advanced Grant MC@NNLO (340983), 
Department of Energy under Contracts No. DE-SC0011090 and DE-AC52-06NA25396, by the Simons Foundation Investigator Grant No. 327942, by the LANL/LDRD Program, within the framework of the TMD Topical Collaboration, and by a startup grant from Zhejiang University.

\bibliography{Higgs_pt_matched_spectrum_bib}

\end{document}